\documentclass[aps,pra,reprint,groupedaddress]{revtex4-1}
\usepackage{dcolumn}
\usepackage{graphicx}
\usepackage{amsmath}
\usepackage{amsfonts}
\usepackage{color}

\begin{document}


\title{Gutzwiller Monte Carlo approach for a critical dissipative spin model}


\author{Wim Casteels$^1$}
\author{Ryan M.~Wilson$^2$}
\author{Michiel Wouters$^1$}
\affiliation{$^1$TQC, Universiteit Antwerpen, Universiteitsplein 1, B-2610 Antwerpen, Belgium} 
\affiliation{$^2$Department of Physics, The United States Naval Academy, Annapolis, MD 21402, USA}


\date{\today}

\begin{abstract}
We use the Gutzwiller Monte Carlo approach to simulate the dissipative XYZ-model in the vicinity of a dissipative phase transition. This approach captures classical spatial correlations together with the full on-site quantum behavior, while neglecting non-local quantum effects. By considering finite two-dimensional lattices of various sizes, we identify a ferromagnetic and two paramagnetic phases, in agreement with earlier studies. The greatly reduced numerical complexity the Gutzwiller Monte Carlo approach facilitates efficient simulation of relatively large lattice sizes.  The inclusion of the spatial correlations allows to describe critical behavior which is completely missed by the widely applied Gutzwiller decoupling of the density matrix.
\end{abstract}


\maketitle

\section{Introduction}

In recent years there has been a growing interest in the quantum behavior of dissipative systems (see for example the reviews \cite{RevModPhys.85.299, 0034-4885-80-1-016401, 2040-8986-18-10-104005}), which can be realized with experimental platforms including ultracold atoms, trapped ions, superconducting circuits, semiconductor microcavities, and microwave cavities.  Correspondingly, there is a renewed interest in dissipative phase transitions and critical behavior in these systems~\cite{PhysRevA.86.012116, PhysRevX.5.031028, PhysRevA.93.033824, PhysRevA.93.023821, PhysRevA.95.012128, PhysRevA.94.033841, 2017arXiv170502865S, 2017arXiv170302934M}, which were recently observed experimentally~\cite{hamel2015spontaneous, PhysRevX.7.011016, PhysRevX.7.011012, PhysRevLett.118.247402, 2017arXiv170701837F}. The numerical description of these critical phenomena is notoriously difficult, due to various factors including the exponentially large Hilbert space, the large entropy, and the critical slowing down of the dynamics. This has motivated various theoretical efforts towards the development of efficient simulation techniques for dissipative systems (see for example Refs. \cite{PhysRevB.78.155117, PhysRevLett.110.233601, PhysRevB.89.245108, PhysRevLett.114.220601, PhysRevA.92.022116, PhysRevLett.115.080604, PhysRevX.6.021037, 1367-2630-18-9-093007}).

The dissipative XYZ-model is a spin model on a lattice described by the anisotropic XYZ-Heisenberg Hamiltonian and a spontaneous decay along the $z$-direction. A very rich mean-field phase diagram was predicted for this system in Ref.~\cite{PhysRevLett.110.257204}. In Ref.~\cite{PhysRevX.6.031011}, short-range correlations were included in the description, which changed the topology of the phase diagram--a remarkable feature not observed for a system at equilibrium. In particular, they identified and examined a ferromagnetic to paramagnetic phase transition, which was not captured by the mean-field approach of Ref. \cite{PhysRevLett.110.257204}. More recently, the susceptibility, the entropy, and the Fisher information were examined with the corner-space renormalization method for 2D lattices with a system size of up to $6 \times 6$ sites \cite{PhysRevB.95.134431}. Possible experimental realizations of the dissipative XYZ-model with trapped ions or ultracold atoms are discussed in Ref. \cite{PhysRevLett.110.257204}. Therein, they show that the model parameters can be externally tuned, which in principle can provide experimental access to interesting regions of the phase diagram.

In this paper, we apply the Gutzwiller Monte Carlo approach to the dissipative XYZ-model to examine the ferromagnetic to paramagnetic dissipative phase transition. This approximation neglects non-local quantum correlations, while capturing the on-site correlations and classical spatial (non-local) correlations. Further, this approach introduces a reduced Hilbert space that scales linearly with system size, allowing for the simulation of relatively large lattices.   By considering individual trajectories, we identify features of the different phases, and for finite lattices sizes we identify an intermediate regime where both phases are metastable and the dynamics exhibit switching between them. The transition is also explored by calculating the spin structure factor which is qualitatively in agreement with the earlier results of Ref.~\cite{PhysRevX.6.031011}. We also examine the spatial dependence of the correlation function which in the paramagnetic phase reveals remnants of a long-range anti-ferromagnetic order. Our results show that it is sufficient to consider the non-local classical fluctuations to at least qualitatively capture the critical behavior of this phase transition, which is not identified by standard Gutzwiller mean-field theories.

The paper is organized as follows: in the second section we introduce the dissipative XYZ-model, the third section discusses the Gutzwiller Monte Carlo approach, and in the fourth section the results of the simulations are presented. Finally, in the last section, the conclusions and perspectives are presented.

\section{The dissipative XYZ-model}
The anisotropic XYZ-Heisenberg Hamiltonian is given by (in units with $\hbar = 1$):
\begin{equation}
\hat{H} = \sum_{<\bold{i},\bold{j}>}\left(J_x \hat{\sigma}_\bold{i}^{(x)}\hat{\sigma}_\bold{j}^{(x)} + J_y\hat{\sigma}_\bold{i}^{(y)}\hat{\sigma}_\bold{j}^{(y)} + J_z \hat{\sigma}_\bold{i}^{(z)}\hat{\sigma}_\bold{j}^{(z)}\right),
\label{Eq: SysHam}
\end{equation} 
where $\sigma_\bold{i}^{(\alpha)}$ are the Pauli matrices (with $\alpha = \{x,y,z\}$) for spin $\bold{i}$ and the sum is over all nearest neighbors. The parameters $J_x$, $J_y$ and $J_z$ are the coupling strengths for the $x$, $y$ and $z$ spin components, respectively. The Hamiltonian (\ref{Eq: SysHam}) determines the unitary part of the time evolution. The full time evolution of the dissipative XYZ-model, including the dissipation along the $z$-direction, is described by a Lindblad-master equation for the density matrix $\hat{\rho}$:
\begin{equation}
\partial_t\hat{\rho} = -i\left[\hat{H},\hat{\rho}\right] + \frac{\gamma}{2} \sum_\bold{j}\left(2\hat{\sigma}_\bold{j}^{(-)}\hat{\rho}\hat{\sigma}_\bold{j}^{(+)}- \left\{\hat{\sigma}_\bold{j}^{(+)}\hat{\sigma}_\bold{j}^{(-)},\hat{\rho}\right\} \right).
\label{Eq: MasEq}
\end{equation} 
Where $\gamma$ is the decay rate of the spins and $\sigma_i^{(\pm)}$ are the raising (+) and lowering (-) operators along the $z$-axis. The master equation (\ref{Eq: MasEq}) has a $\mathbb{Z}_2$ symmetry: $(\hat{\sigma}_n^{(x)},\hat{\sigma}_n^{(y)}) \to (-\hat{\sigma}_n^{(x)},-\hat{\sigma}_n^{(y)})$. This symmetry can be spontaneously broken in an ordered phase such as a ferromagnet with a finite magnetization in the $xy$-plane \cite{PhysRevLett.110.257204, PhysRevX.6.031011}. 

The dissipative XYZ-model has been studied by various methods. In Ref. \cite{PhysRevLett.110.257204} a single-site mean-field calculation revealed that the competition between precessional and dissipative dynamics leads to a very rich phase diagram. In Ref. \cite{PhysRevX.6.031011} more advanced numerical approaches were applied to a particular region of the phase diagram containing a paramagnetic to ferromagnetic phase transition. Using a matrix product operator Ansatz for the density matrix they revealed the absence of a phase transition in 1 dimension. For 2 dimensions they included the non-local correlations on a relatively short range with the cluster mean-field approach. This revealed that while the single-site mean-field approach predicts a ferromagnetic phase all the way to $J_y \to \infty$, the cluster mean-field approach predicts a paramagnetic phase for large $J_y$. For the parameter values $J_x = 0.9\gamma$ and $J_z = \gamma$ a finite size scaling revealed that in the thermodynamic limit the system exhibits a ferromagnetic phase for $1.04\gamma \lesssim J_y\lesssim 1.4\gamma$ \cite{PhysRevX.6.031011}. We will consider the same parameter range.

\section{The Gutzwiller Monte Carlo approach}
The Gutzwiller Monte Carlo approach is a combination of the Gutzwiller Ansatz for the wave function and the wave function Monte Carlo simulation. We start by briefly discussing these two methods. 

The Gutzwiller Ansatz $\Psi_{GW}$ corresponds to a product wave function \cite{PhysRevLett.10.159}:
\begin{equation}
\Psi_{GW}\left(\{i\} \right) = \prod_i{\psi_i},
\label{Eq: GW}
\end{equation} 
where the index $i$ denotes the different modes or lattice sites and $\psi_i$ is a single mode wave function on site $i$. This Ansatz neglects the spatial correlations while the local correlations are fully taken into account. The dimension of the wavefunction $\Psi_{GW}$ grows linearly with the system size $N$ as $Nd$, with $d$ the local Hilbert space dimension, rather the exponential increase $d^N$ for the full wave function. A well-known application of the Gutzwiller wavefunction (\ref{Eq: GW}) is for the description of a superfluid to Mott insulator transition \cite{PhysRevB.44.10328, PhysRevB.45.3137}. More recently, the Gutzwiller Ansatz has been extended for the description of dissipative lattice systems by decoupling the full density matrix as a direct product of single-site density matrices (see for example Refs. \cite{PhysRevLett.105.015702, PhysRevLett.110.257204, PhysRevX.6.031011, PhysRevA.90.023827, PhysRevLett.110.233601, PhysRevA.94.033801, 2016arXiv161100697B}). 

The wave function Monte Carlo (also known as quantum trajectories) is a simulation technique for dissipative systems which keeps track of the wave function during individual realizations \cite{PhysRevLett.68.580, Molmer:93, PhysRevLett.70.2273}. The evolution of the wave function is determined by stochastically simulating an external measurement of the excitations that leave the system. This corresponds to an exact simulation technique and the full density matrix can be obtained by averaging over the individual realizations. 

If one is interested in describing systems containing multiple coupled modes, such as a lattice model, one quickly runs into a prohibitively large Hilbert space which grows exponentially with the system size. The Gutzwiller Monte Carlo approach deals with this by combining the relative small effective Hilbert space of the Gutzwiller Ansatz (\ref{Eq: GW}) with the wave function Monte Carlo. In practice, a wave function Monte Carlo simulation is performed while restricting the wave function to the subspace of product wave functions (\ref{Eq: GW}). This captures all the local correlations while non-local quantum correlations are neglected. Importantly, non-local classical correlations are included by averaging over the different realizations. These are neglected with a Gutzwiller decoupling of the density matrix. Recently this approach was applied for the description of the optically bistable driven-dissipative Bose-Hubbard dimer \cite{PhysRevA.95.043833}. 

The validity of the method is determined by the role of the non-local quantum correlations and whether classical correlations are sufficient to describe the system. This argument could be reversed by noting that if the Gutzwiller Monte Carlo approach succeeds in capturing the behavior of the system non-local quantum correlations are not important. An interesting question is then to which extent the critical behavior of dissipative systems is captured. This is also motivated by recent works that show that some critical dissipative quantum models can be mapped onto a classical system \cite{PhysRevB.93.014307, PhysRevA.95.043826}. With this in mind we explore the properties of the critical dissipative XYZ-model with the Gutzwiller Monte Carlo approach.

\section{Results}
For the simulations we consider finite 2D symmetric lattices of various sizes with periodic boundary conditions and total number of lattice sites $N$. We always use the system parameters $J_x = 0.9\gamma$ and $J_z = \gamma$ and vary $J_y$. For these parameters the single-site mean-field approach predicts a transition from a paramagnetic to a ferromagnetic phase at $J_y \approx 1.04\gamma$ \cite{PhysRevLett.110.257204}. The more advanced cluster mean-field approach revealed another ferromagnetic to paramagnetic phase transition at $J_y \approx 1.4\gamma$ in the thermodynamic limit \cite{PhysRevX.6.031011}.  

Care has to be taken when choosing the initial state for the Gutzwiller Monte Carlo simulations. A possible issue arises when the system reaches the pure state with all spins pointing downwards in the $z$-direction. This is the exact steady-state of the dissipative XXZ-model, corresponding to $J_x = J_y$ \cite{PhysRevLett.110.257204}. However, within the considered Gutzwiller approximation, it becomes an artificial dark state for all possible system parameters since it is not possible to reach a finite magnetization in the $xy$-plane starting from zero. The same behavior is encountered with the single-site mean field approach \cite{PhysRevLett.110.257204}. This is clearly an artifact of the approximation since the unitary evolution in the full Hilbert space, determined by the XYZ-Hamiltonian (\ref{Eq: SysHam}), can lead to fluctuations of the magnetization in the $xy$-plane starting from zero (this can easily be verified by just considering two coupled spins). Because of this the initial state should have a finite magnetization in the $xy$-plane and in practice we always consider an initial state with all spins aligned with the $x$-direction. 

\begin{figure}[h]
  \includegraphics[scale=0.5]{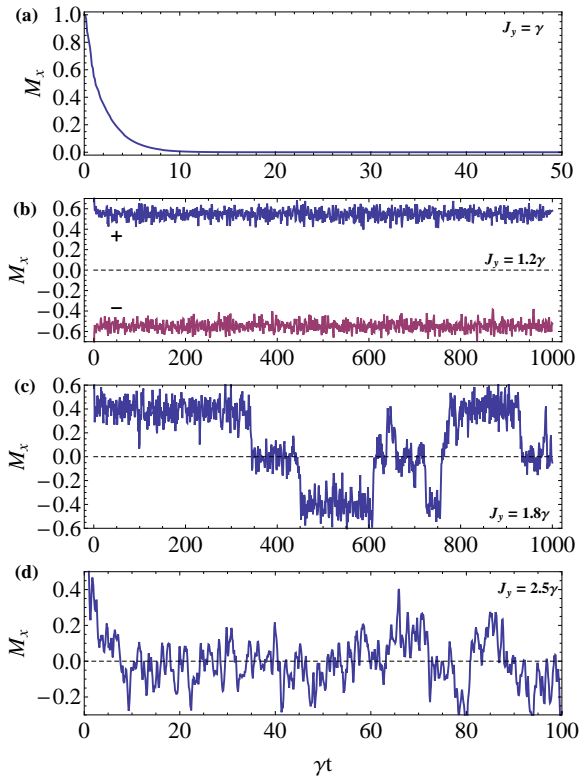}
  \caption{\label{Fig2} The magnetization in the $x$-direction (\ref{Eq: Mx}) as a function of time on an $6 \times 6$ lattice during individual trajectories for 4 representative values of $J_y$: $\gamma$ (a), $1.2\gamma$ (b), $1.8\gamma$ (c) and $2.5\gamma$ (d). Initially all spins are aligned parallel with the x-axis in the positive direction (except for the second panel with $J_y = 1.2$ where results are shown with initial spins both in the positive (+) and in the negative (-) $x$-direction). The different phases are clearly visible: a paramagnetic phase with $M_x=0$ for $J_y = \gamma$, a ferromagnetic phase with $M_x \ne 0$ for $J_y = 1.2\gamma$ and another paramagnetic phase for $J_y = 2.5\gamma$ with $M_x=0$ on average. For $J_y = 1.8\gamma$ the system exhibits an intermediate regime where the system switches between the different phases on a relatively large timescale (note the different scales for the time). }
\end{figure}  

We start by considering the magnetization $M_x$ in the $x$-direction:
\begin{equation}
M_x = N^{-1}\sum_\bold{i} \langle \hat{\sigma}_\bold{i}^{(x)}\rangle.
\label{Eq: Mx}
\end{equation} 
This quantity is presented in Fig. \ref{Fig2} for an $6 \times 6$ lattice as a function of time during single trajectories. The different panels correspond to different values of $J_y$. For $J_y = \gamma$ (a) the system relaxes to a pure steady-state with all spins pointing downward in the $z$-direction and $M_x = 0$, as also predicted by the single-site mean-field approach \cite{PhysRevLett.110.257204}. As the system gets trapped in this artificial dark state (see discussion above) the fluctuations around this state, as observed in Refs. \cite{PhysRevX.6.031011, PhysRevB.95.134431}, are not captured. For $J_y = 1.2\gamma$ (b) the system exhibits a ferromagnetic phase with on average a non-zero magnetization. Depending on whether the initial spins are aligned with the positive (+) or negative (-) $x$-direction $M_x$ stays positive or negative, respectively. For the value $J_y = 1.8\gamma$ (c) the system switches on a relatively long timescale between the two ferromagnetic phases which are stable for smaller values of $J_y$ and a paramagnetic phase which becomes stable at larger $J_y$. This paramagnetic phase is clearly visible for $J_y = 2.5 \gamma$ (d) with a magnetization which fluctuates around zero.

\begin{figure}[h]
  \includegraphics[scale=0.5]{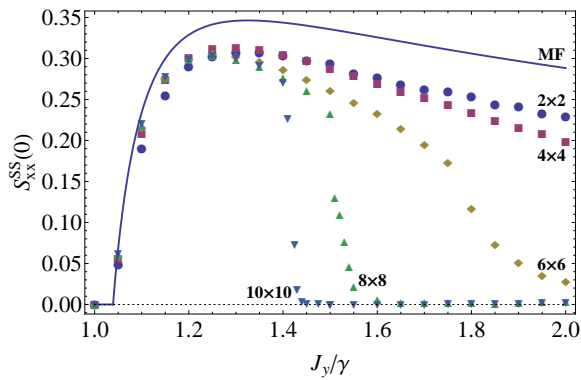}
  \caption{\label{Fig1} The spin structure factor $S^{xx}_{SS}(0)$ (\ref{Eq: Cor}) for finite two-dimensional lattices as a function of the coupling $J_y$ (in units of $\gamma$). The full line is the single-site mean-field (MF) result (\ref{Eq: CorMF}) which predicts a non-zero value for $J_y \gtrsim 1.04$, corresponding to a ferromagnetic phase. The markers are the results from the Gutzwiller Monte Carlo approach obtained by averaging a single trajectory over a total time of $10.000/ \gamma$ for various system sizes (as indicated in the figure) and $40.000/\gamma$ in the intermediate regime for the two largest lattices. This reveals a transition to a paramagnetic phase with $S^{xx}_{SS}(0) = 0$ for large $J_y$ which as the system size is increased becomes sharper with a shrinking ferromagnetic region.}
\end{figure}  

To further explore the different phases in the steady-state we now consider the steady-state spin structure factor $S^{xx}_{SS}(\bold{k}=\bold{0})$, where:
\begin{equation}
S^{xx}_{SS}(\bold{k}) = \frac{1}{N (N-1)}\sum_{\bold{j} \neq \bold{l}}e^{-i \bold{k}.(\bold{j}-\bold{l})}\langle\hat{\sigma}_\bold{j}^{(x)}\hat{\sigma}_\bold{l}^{(x)}\rangle.
\label{Eq: Cor}
\end{equation} 
A non-zero value of $S^{xx}_{SS}(\bold{0})$ indicates the presence of a ferromagnetic phase and it is zero in the paramagnetic phase. In Fig. \ref{Fig1} $S^{xx}_{SS}(\bold{0})$ is presented as a function of $J_y$ for different lattice sizes. The steady-state results are obtained by averaging a single trajectory over a total time of $10.000/\gamma$ and $40.000/\gamma$ in the intermediate regime for the two largest lattices. The result from the single-site mean-field (MF) approach is also presented which predicts a ferromagnetic phase for $J_y \gtrsim 1.04$, with:
\begin{eqnarray}
S^{xx}_{SS}(\bold{0}) = && M_x^{(MF)}M_x^{(MF)} \nonumber\\
= &&\frac{\gamma}{8}\left(\frac{\gamma}{16}\sqrt{\frac{1}{(J_z - J_x)(J_y - J_z)}} -1\right) \nonumber \\
&&\times \sqrt{\frac{1}{(J_z - J_x)(J_y - J_z)}} \frac{J_y - J_z}{J_x - J_y},
\label{Eq: CorMF}
\end{eqnarray}  
where $M_x^{(MF)}$ is the single-site mean-field result for the magnetization in the $x$-direction \cite{PhysRevLett.110.257204, PhysRevX.6.031011}. Close to the transition at $J_y \approx 1.04$ the results for finite lattice sizes in Fig. \ref{Fig1} are in good agreement with the single-site mean-field result (\ref{Eq: CorMF}) which is also predicted by the cluster mean-field approach \cite{PhysRevX.6.031011}. For values of $J_y$ just above this transition the results converge to a finite value as a function of the lattice size, revealing the presence of a ferromagnetic phase in the thermodynamic limit. For large values of $J_y$ the spin structure factor $S^{xx}_{SS}(\bold{0})$ decreases to zero as the lattice size is increased, indicating a paramagnetic phase in the thermodynamic limit. These results show the presence of a ferromagnetic to paramagnetic transition in the thermodynamic limit which is not captured by the single-site mean-field result (\ref{Eq: CorMF}), as also predicted by the cluster mean-field approach \cite{PhysRevX.6.031011}. As the system size is increased the ferromagnetic region shrinks and the transition becomes sharper which indicates that in the thermodynamic limit the intermediate regime where both phases are metastable (see third panel of Fig. \ref{Fig2}) disappears. The observed behavior is in qualitative agreement with the results of the cluster mean-field approach which predicts a ferromagnetic phase for $1.04\gamma \lesssim J_y \lesssim 1.4\gamma$ in the thermodynamic limit \cite{PhysRevX.6.031011}.

\begin{figure}[h]
  \includegraphics[scale=0.55]{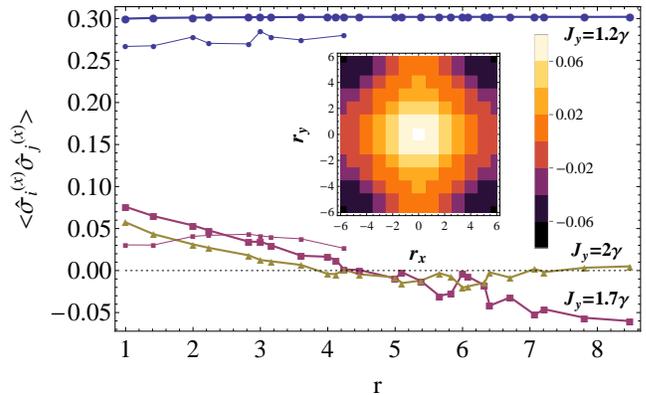}
  \caption{\label{Fig3} The correlation function $\langle\hat{\sigma}_\bold{i}^{(x)}\hat{\sigma}_\bold{j}^{(x)}\rangle$ as a function of the distance $r = |\bold{i}-\bold{j}|$ for a $12 \times 12$ lattice and for $J_y = 1.2\gamma$ (circles), $1.7 \gamma$ (squares) and $2\gamma$ (triangles). The results are obtained with the Gutzwiller Monte Carlo approach by averaging a single trajectory over a total time of $20.000/ \gamma$. For $J_y = 1.2\gamma$ the correlation function converges to a finite value as expected in the ferromagnetic phase. For $J_y = 1.7\gamma$ the correlation function decays and becomes negative at relatively large distances, corresponding to a long-scale anti-ferromagnetic ordering. For larger distances the correlation function is expected to eventually decay to zero, as observed for $J_y = 2\gamma$. The smaller symbols are the result from the cluster mean-field approach for $J_y = 1.2 \gamma$ and $1.7 \gamma$ \cite{PhysRevX.6.031011}. The inset shows the correlation function as a function of the distance in the $x$- and $y$-direction for $J_y = 1.7 \gamma$.}
\end{figure} 

In Fig. \ref{Fig3} the average steady-state correlation function $\langle\hat{\sigma}_{\bold{i}}^{(x)}\hat{\sigma}_{\bold{j}}^{(x)}\rangle$ is presented for a $12 \times 12$ lattice as a function of the Euclidian distance $r = |\bold{i}-\bold{j}|$. For $J_y = 1.2\gamma$ the correlation function converges to a non-zero value for large distances, as expected for the ferromagnetic phase. For $J_y = 1.7\gamma$ the correlation function decays and becomes negative at relatively large distances, revealing remnants of anti-ferromagnetic ordering at a relatively long scale. For larger distances the correlation function is expected to decay towards zero, as observed for $J_y = 2\gamma$ and as expected for the paramagnetic phase.  As a consequence of the periodic boundary conditions not all the points in Fig. \ref{Fig3} are on a smooth curve. To clarify this, the correlation function is also plotted as a function of the distance in the $x$- and in the $y$-direction in the inset of Fig. \ref{Fig3}. Along a lattice direction the boundary is reached at a shorter distance with respect to a diagonal direction. For larger lattices this effect should become weaker and in the thermodynamic limit the points are expected to converge to a smooth curve. The results from the cluster mean-field approach for a $4 \times 4$ cluster of Ref. \cite{PhysRevX.6.031011} are also presented in Fig. \ref{Fig3} for $J_y = 1.2\gamma$ and $1.7 \gamma$ (smaller symbols). For $J_y = 1.2\gamma$ these results qualitatively agree but our correlation is a bit stronger. For $J_y = 1.7\gamma$ the correlation function obtained in Ref. \cite{PhysRevX.6.031011}initially increases and then weakly decreases which is different from the Gutzwiller Monte Carlo prediction which exhibits a monotoneous decaying behavior at this length scale. The length scale considered with the cluster mean-field approach is too small to compare the long-range ferromagnetic behavior. It is not clear whether these deviations are due to the relatively small size of the considered cluster in Ref. \cite{PhysRevX.6.031011} or due to the Gutzwiller approximation. It would be interesting to compare the results with other numerical approaches that capture all correlations such as the corner-space renormalization method \cite{PhysRevLett.115.080604, PhysRevB.95.134431}. 

\section{Conclusions and perspectives}
We considered the Gutzwiller Monte Carlo approach for the description of the critical dissipative XYZ-model which allows to efficiently simulate relatively large lattice sizes. This corresponds to neglecting the non-local quantum correlations while capturing the classical spatial correlations. Our results reveal the presence of a transition from a ferromagnetic to paramagnetic phase which is not captured by the single-site mean field treatment presented in Ref. \cite{PhysRevLett.110.257204}. This behavior is at least qualitatively in agreement with earlier work that includes short-range correlations in Ref. \cite{PhysRevX.6.031011}. A comparison of the spatial correlation function on the other hand revealed deviations with the results of Ref. \cite{PhysRevX.6.031011} which could be better understood by comparing with a more advanced numerical approach. This reveals the potential of the Gutzwiller Monte Carlo approach to become an important tool for the description of dissipative phase transitions.

The approach can be straightforwardly generalized to other dissipative lattice systems such as the driven-dissipative Bose-Hubbard model \cite{AMO2016934} and a lattice of coupled Jaynes-Cummings resonators \cite{PhysRevX.7.011016}. This could shed new light on the role of quantum correlations for dissipative phase transitions. Also dynamical properties of critical systems could be explored by the Gutzwiller Monte Carlo approach. This gives access to the Liouvillian gap which has been shown to be an important property for the characterization of dissipative phase transitions \cite{PhysRevA.86.012116, PhysRevA.94.033801, PhysRevA.95.012128, PhysRevX.7.011016, 2017arXiv170701837F}.

In the same spirit as the Gutzwiller Monte Carlo approach one could use other variational wavefunctions in combination with a wave function Monte Carlo simulation. 
A straightforward extension would be to include short-range quantum correlations in the wavefunction. This could be done by considering a sublattice of clusters and keeping track of all (quantum and classical) correlations within the clusters while only classical correlations are considered between them. For the considered dissipative XYZ-model this could solve the issue with the artificial dark state already by considering a sublattice of clusters of size $2 \times 1$.

\acknowledgements{This work was financially supported by the FWO Odysseus program.  RMW acknowledges partial support from the National Science Foundation under Grant No. PHY-1516421.}  

\bibliography{manusc}

\end{document}